# A Dynamic Subgrid-Scale Model Based on Liutex Theory for Wall-Bounded Turbulent Flows

Jiawei Chen[a], Yifei Yu[b], Chaoqun Liu[b]*

[a]Department of Aerospace Engineering, Iowa State University, Ames, Iowa 50011, USA

[b]Department of Mathematics, University of Texas at Arlington, Arlington, Texas 76019, USA

* Correspondence: cliu@uta.edu

**Abstract:** Accurate subgrid-scale (SGS) modeling remains a major challenge in large eddy simulation (LES), particularly for wall-bounded turbulent flows with strong near-wall anisotropy. This study proposes a novel SGS model based on Liutex theory, featuring a dynamically adaptive model coefficient and a physics-based length-scale formulation. The magnitude of the Liutex vector is employed as the characteristic velocity scale, enabling a direct and objective quantification of local vortical intensity. The length scale is determined from local flow properties and reflects the physical nature of turbulent diffusion, which occurs predominantly in directions perpendicular to the rotation axis. The dynamic model coefficient adapts locally to variations in vortical structures and naturally vanishes at the wall. Importantly, this coefficient is free of empirical tuning, as it is derived rigorously from the triple decomposition. The model is validated through LES of turbulent channel flow at the friction Reynolds number of 544. Results show improved predictions of both normal and shear Reynolds stresses, particularly in near-wall regions, compared with other SGS models. The proposed model demonstrates competitive computational efficiency, with only a 4.21% overhead.

**Keywords:** subgrid-scale modelling, Liutex, model coefficient, length scale, computational efficiency

## 1. Introduction

With the rapid advancement of computational power in recent years, large eddy simulation (LES) has gained widespread adoption. In terms of computational cost, LES falls between Reynolds-averaged Navier–Stokes (RANS) and direct numerical simulation (DNS) [1]. LES provides a more accurate simulation of unsteady flow dynamics by resolving large-scale turbulent structures and modeling the smaller ones.

The subgrid-scale (SGS) stress model accounts for the scales smaller than the resolved scales and plays a crucial role in LES. In 1963, Smagorinsky[2] proposed the earliest SGS stress model, which employs the resolved strain-rate tensor to define the velocity scale. While it performs well in decaying homogeneous isotropic turbulence[3][4], it does not vanish in laminar and near-wall regions, limiting its ability to capture transitional and wall-bounded flows accurately. To overcome this, the Smagorinsky model is often combined with damping functions[5][6]. Lilly[7] estimated the model coefficient as 0.17 for isotropic turbulence in the inertial subrange, while Deardorff[8] suggested 0.1 for wall-bounded turbulent shear flows. In 1999, Nicoud[9] proposed the Wall-Adapting Local Eddy-viscosity (WALE) model, which defines the velocity scale using both the strain and rotation. A key advantage of the WALE model is its natural vanishing of eddy viscosity at the wall and its favorable near-wall scaling behavior of $O(y^3)$. In 2004, Vreman[10] proposed a SGS model in which the velocity scale is constructed from a specific combination of the velocity gradient tensor. This model uses first-order derivatives and avoids the need for explicit filtering, averaging, or clipping. It is also rotationally invariant under isotropic filtering. More recently, Ding et al.[11] (2022) proposed a novel SGS model employing the rotational part of the velocity gradient tensor to define the velocity scale. A key advantage is that its eddy viscosity naturally decays to zero within the viscous sublayer of wall-bounded flows.

Dynamic SGS models are often employed to resolve the issue of determining model coefficients in subgrid-scale (SGS) models. Germano et al. [12] developed the dynamic Smagorinsky model, where the model coefficient is determined based on the local flow conditions rather than using a fixed constant. It effectively dissipates energy from large scales in isotropic decaying turbulence and appropriately reduces to zero in laminar and transitional flows[13]. Lilly[14] modified the subgrid-scale closure method developed by Germano et al. [12] using a least-squares technique to minimize the discrepancy between the closure assumption and the resolved stresses. Like the Smagorinsky model, the WALE model also lacks a universal coefficient and requires case-specific calibration, particularly for complex geometries. To overcome this limitation, Toda et al.[15] incorporated the Germano–Lilly procedure into the WALE model, leading to the development of the dynamic WALE model.

This paper proposes a novel subgrid-scale (SGS) model grounded in Liutex theory. The characteristic velocity scale is defined by the magnitude of the Liutex vector to directly represent local vortical intensity, while the length scale is constructed to account for the turbulent diffusion perpendicular to local rotation axes. A dynamic procedure is introduced to compute the model coefficient locally based on the relative strength of vortical structures, ensuring physical consistency and adaptability across different flow conditions. The performance of the proposed model is systematically evaluated through large eddy simulations of turbulent channel flow at a friction

Reynolds number of $Re_\tau = 544$. Detailed analyses are conducted to assess Reynolds stress predictions, near-wall behavior, and the underlying physical mechanisms, with comprehensive comparisons against the Dynamic Smagorinsky model and the model of Ding et al.[11]. Computational efficiency and grid sensitivity are also examined to demonstrate the robustness and practical applicability of the proposed SGS model.

The remainder of this paper is organized as follows. Section 2 introduces the governing equations, subgrid-scale models, and numerical methods. Section 3 validates the model capabilities using the turbulent channel flow. Finally, Section 4 provides the conclusions of this study.

## 2. Methodology

This section presents the theoretical framework and numerical implementation of the proposed Liutex-based subgrid-scale model for large eddy simulation of turbulent flows. The methodology is organized into three principal components. First, the filtered Navier-Stokes equations and the standard eddy-viscosity formulation for SGS stress modeling are introduced in Section 2.1, establishing the mathematical foundation for LES. Section 2.2 details the novel SGS model formulation, with particular emphasis on three key innovations: (A) the characteristic velocity scale based on the Liutex vector magnitude, (B) the physics-informed length scale that accounts for diffusion perpendicular to rotation axes, and (C) the dynamic model coefficient computed from local flow characteristics. Finally, Section 2.3 describes the numerical methods employed for spatial discretization and temporal integration within the OpenFOAM framework.

### 2.1 Governing equations and subgrid-scale stress models

First, the grid filter (denoted by overbar) is applied to the Navier-Stokes equations, resulting in

$$\frac{\partial \overline{v}_i}{\partial t} + \frac{\partial}{\partial x_j}\left(\overline{v}_i \overline{v}_j\right) = -\frac{1}{\rho}\frac{\partial \overline{p}}{\partial x_i} + \nu \frac{\partial^2 \overline{v}_i}{\partial x_j x_j} - \frac{\partial \tau_{ij}}{\partial x_j} \quad (1)$$

Where the SGS stress tensor $\tau_{ij}$ is defined by

$$\tau_{ij} = \overline{v_i v_j} - \overline{v}_i \overline{v}_j \quad (2)$$

The linear eddy-viscosity model is used to relate the anisotropic SGS stress to the filtered rate of strain $\bar{S}_{ij}$.

$$\tau_{ij} - \frac{1}{3}\delta_{ij}\tau_{kk} = -2\nu_t \bar{S}_{ij} \tag{3}$$

## 2.2 Current subgrid-scale model

This section provides a detailed description of the proposed SGS model, with particular emphasis on the physical interpretation of the velocity scale, length scale, and model coefficient. The characteristic velocity scale is defined by the magnitude of the Liutex vector to directly represent local vortical intensity, while the length scale is constructed to account for the anisotropic nature of turbulent diffusion perpendicular to local rotation axes. A dynamic procedure is introduced to compute the model coefficient locally based on the relative strength of vortical structures, ensuring physical consistency and adaptability across different flow conditions.

By analogy to the mixing-length hypothesis, the eddy viscosity is modelled as

$$\nu_t = l_R^2 \bar{R} = \left(C_R \Delta_R\right)^2 R \tag{4}$$

where $\bar{R}$ denotes the characteristic filtered magnitude of the Liutex vector (hereinafter simplified as R for brevity), and $l_R$ is a length scale assumed proportional to the filter width $\Delta_R$ via the model coefficient $C_R$.

### A. Characteristic velocity scale

The velocity-scale formulation is inspired by the work of Ding et al.[11], who employed the magnitude of the Liutex vector[16] **R** as the characteristic velocity scale in the eddy-viscosity model. Subsequently, Chen et al.[17] further investigated the model's near-wall scaling behavior and its capability to predict flow separation. Hossen et al.[18] applied the model of Ding et al.[11] to canonical flows, including backward-facing step flow, channel flow, and flat-plate boundary-layer transition. **R** characterizes the local rigid-body rotation of fluid and was originally developed for vortex identification and analysis[19][20][21]. Unlike other vortex identification methods, such as Q[22], $\lambda_2$[23], $\Delta$[24] and $\lambda_{ci}$[25], **R** is a vector. Its direction represents the axis of rotation, while its magnitude is twice the local angular velocity of rigid rotation. Kolář and Šístek[26] showed that $\boldsymbol{R}$ is unaffected by stretching or shear, confirming its robustness in vortex identification. The explicit formula[27] of $\boldsymbol{R}$ vector is given by

$$\boldsymbol{R} = R\boldsymbol{r} = \left[ \boldsymbol{\omega} \cdot \boldsymbol{r} - \sqrt{(\boldsymbol{\omega} \cdot \boldsymbol{r})^2 - 4\lambda_{ci}^2} \right] \boldsymbol{r} \qquad (5)$$

Where $\boldsymbol{\omega}$ is the vorticity vector, $\lambda_{ci}$ is the imaginary part of the complex conjugate eigenvalue of the velocity gradient tensor, and $\boldsymbol{r}$ is the eigenvector corresponding to the real eigenvalue.

## B. Sugird-scale Length

Conventional length scales employed in SGS models, such as the maximum cell dimension $\Delta_{max}$ (Equation 6) and the cube root of cell volume $\Delta_{vol}$ (Equation 7), rely exclusively on local grid metrics without consideration of flow characteristics:

$$\Delta_{max} = \max(\Delta_x, \Delta_y, \Delta_z) \qquad (6)$$

$$\Delta_{vol} = \sqrt[3]{\Delta_x \Delta_y \Delta_z} \qquad (7)$$

where $\Delta_x$, $\Delta_y$ and $\Delta_z$ represent the cell spacing in the x, y, and z directions, respectively.

In contrast, the present formulation adopts a flow-based length scale that adapts to local turbulent structures. Recognizing that turbulent dissipation occurs predominantly in directions perpendicular to the rotation axis, we define the length scale as the characteristic dimension perpendicular to the local vortex orientation. This length scale, denoted as $\Delta_R$, is computed using:

$$\Delta_R = 2 \max_{m=1,6} \left| \mathbf{n}_R \times \mathbf{r}_{cf} \right| \qquad (8)$$

Where $\boldsymbol{n}_\omega$ is the unit vector in the vorticity direction, $\boldsymbol{n}_R$ is the unit Liutex vector. $\boldsymbol{r}_{cf}$ denotes the vector connecting the cell center to each face center.

Figure 1 shows the geometric illustration of the Liutex-based length scale $\Delta_R$ computation. Equation (8) quantifies twice the maximum magnitude of the cross product between the unit Liutex vector and the cell center-to-face vectors. Geometrically, this operation identifies the maximum characteristic length perpendicular to the vortical axis within each computational cell. This approach fundamentally differs from conventional grid-based length scales by directly incorporating local flow physics rather than relying solely on cell dimensions.

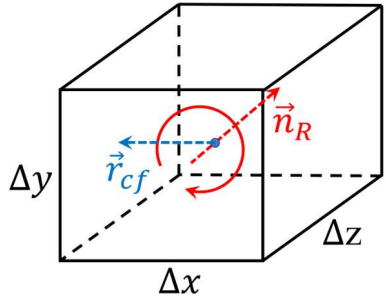

Fig. 1. Geometric illustration of the Liutex-based length scale $\Delta_R$ computation.

## C. Dynamic model coefficient

A key distinguishing feature of the proposed model is the dynamic coefficient $C_R$, which controls the subgrid-scale length scale. The characteristic eddy size is defined as $l_R = C_R \Delta_R$, where $l_R$ represents an equivalent eddy whose turbulent kinetic energy matches the average contribution of the unresolved eddies within a computational cell.

The coefficient $C_R$ is determined dynamically based on the local physical characteristics of the resolved flow, thereby enabling the model to adapt to a wide range of flow conditions. Within each grid cell, turbulent motions span a spectrum of sizes rather than a single characteristic size. To account for the cumulative effect of these multi-scale turbulent structures, a representative average eddy size is introduced, as schematically illustrated in Fig. 2.

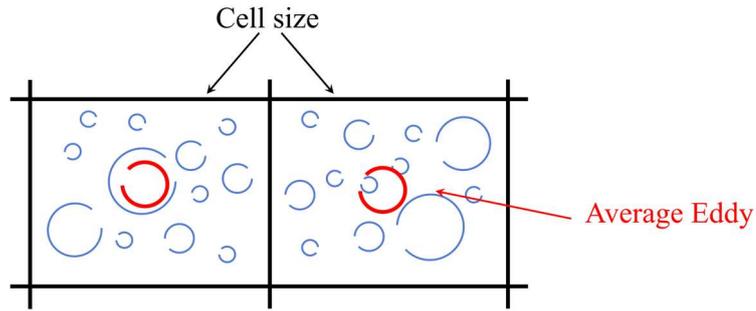

Fig. 2. The representative average eddy size and physical basis of the dynamic model coefficient $C_R$

The relative Liutex is adopted as the dynamic model coefficient $C_R$. It is derived from the triple decomposition[28][29][30] of the velocity gradient tensor $G_{ij}$ (VGT).

Traditionally, the velocity gradient tensor $G_{ij}$ is partitioned into a symmetric strain-rate tensor and an antisymmetric vorticity tensor. In contrast, triple decomposition partitions the VGT into three physically distinct components: normal straining, pure shearing, and rigid-body rotation. The VGT can be expressed in its principal reference frame. In this frame, the VGT is quasi-triangular, and it can be decomposed as:

$$G_{ij} = G_{ij}^N + G_{ij}^R + G_{ij}^S = \begin{bmatrix} \varepsilon_1 & 0 & 0 \\ 0 & \varepsilon_2 & 0 \\ 0 & 0 & \varepsilon_3 \end{bmatrix} + \begin{bmatrix} 0 & 0 & 0 \\ 0 & 0 & \varphi_1 \\ 0 & -\varphi_1 & 0 \end{bmatrix} + \begin{bmatrix} 0 & \gamma_3 & \gamma_2 \\ 0 & 0 & \gamma_1 \\ 0 & 0 & 0 \end{bmatrix} \quad (9)$$

where $G_{ij}^N$, $G_{ij}^R$ and $G_{ij}^S$ denote the normal straining, rigid rotation and pure shearing tensors, respectively. These tensors can be determined and transformed to the original coordinate system using the ordered real Schur decomposition of $G_{ij}$[31].

Correspondingly, the strength of the velocity gradients can be expressed as

$$G_{ij}G_{ij} = G_{ij}^N G_{ij}^N + G_{ij}^R G_{ij}^R + G_{ij}^S G_{ij}^S + G_{ij}^R G_{ij}^S \quad (10)$$

The first three terms represent the strengths of the constituents in Eqn. (15), and the last term represents the interaction between shearing and rigid rotation. Furthermore, the velocity gradient partitioning is defined in terms of the relative contributions of these constituents to $G_{ij}G_{ij}$. The relative contribution of rigid rotation is $gg_{rr} = G_{ij}^R G_{ij}^R / G_{ij}G_{ij}$, which is used as the dynamic model coefficient $C_R$.

The dynamic model coefficient of the present SGS model is defined in Equation (11). The factor of 2 appearing in the denominator of $C_R$ originates from the triple decomposition and is not an empirically introduced constant. Owing to its dynamic formulation, $C_R$ can automatically adapt to local variations in turbulent intensity and flow structure. It should be noted that the dynamic model coefficient $C_R$ vanishes at the wall, since vortical structures, quantified by the Liutex magnitude $\boldsymbol{R}$, do not exist exactly at the wall boundary.

$$C_R = \left( \frac{G_{ij}^R G_{ij}^R}{G_{ij}G_{ij}} \right)^2 = \left( \frac{R^2}{2tr\left[ \left( \nabla \vec{V} \right)^T \left( \nabla \vec{V} \right) \right]} \right)^2 \quad (11)$$

We also briefly review the Germano–Lilly dynamic procedure[12] [14], which was originally derived for the Smagorinsky model to determine the model coefficient in a mathematically consistent manner. When this dynamic procedure is applied to the present SGS model, a corresponding Germano–Lilly dynamic coefficient can be obtained. The resulting expression for the dynamic coefficient $C_D$ is given by Eq. (12),

$$C_D = -\frac{L_{ij}M_{ij}}{2M_{ij}M_{ij}} \qquad (12)$$

where $L_{ij}$ denotes the Leonard stress, representing the additional stress induced by test filtering, and is defined as

$$L_{ij} = \widetilde{\overline{v_i v_j}} - \widetilde{\overline{v}}_i \widetilde{\overline{v}}_j \qquad (13)$$

Here, the overbar denotes the grid-scale filter, while the tilde denotes the test filter. The tensor $M_{ij}$ is defined in Eq. (13) as

$$M_{ij} = \left( \tilde{\Delta}^2 \widetilde{\overline{R}}\widetilde{\overline{S}}_{ij} - \Delta^2 \widetilde{\overline{R}\overline{S}_{ij}} \right) \qquad (14)$$

It is well known that the dynamic coefficient $C_D$ exhibits strong spatial and temporal fluctuations, which may lead to numerical instability. To alleviate this issue, averaging of $C_D$ along homogeneous directions is commonly employed. Moreover, clipping of $C_D$ is required to ensure the positivity of the total viscosity, i.e., $\nu + \nu_{sgs} \geq 0$. For general three-dimensional flows, where no homogeneous direction exists, local averaging combined with coefficient clipping (i.e., constraining $C_D$ within prescribed bounds) is typically adopted.

## 2.3. Numerical methods

In this paper, the open-source software OpenFOAM[32] is used, which is based on the finite volume method for discretizing and solving the governing equations. The flow variables are stored at the centroids of the control volumes (CVs), where the values approximate the local properties of the CV.

In the present simulations, the second-order backward differencing scheme was used for time marching. The face-fluxes of momentum were calculated using a linear interpolation scheme. The same scheme was also used to evaluate the values of the gradients in the centroids of the faces. The face-fluxes of the subgrid scale turbulent kinetic energy were calculated using a TVD interpolation scheme based on upwind and central differencing. The scheme is based on a flux limiter of the form, $\max(\min(2r, 1), 0)$, where $r$ is the ratio of successive gradients.

The solver pimpleFoam was used to solve the equations. The algorithm implemented in the solver is based on a blend of the transient SIMPLE and PISO algorithms, a thorough description of which can be found in Ferziger et al.[33] and Versteeg et al. [34]. The convergence criterion for the numerical solutions is that there is an absolute root-mean-square residual of all equations less than $1.0\times10^{-6}$.

## 3. Results and discussions

Fully developed turbulent channel flow is a theoretical prototype characterized by flow between two infinite parallel planes driven by a constant pressure gradient. It is simulated in a domain of $(L_x, L_y, L_z) = (12.8h, 2.0h, 4.8h)$, with $h$ the channel half-height. Non-slip boundary conditions are imposed in the y direction and periodic boundary conditions are applied to the x and z directions. The bulk Reynolds number is defined as $Re_b = 2u_b h/\nu$, with $u_b$ the bulk velocity and $\nu$ the kinematic viscosity. The friction Reynolds number is defined as $Re_\tau = u_\tau h/\nu$, where $u_\tau = (\tau_w/\rho)^{1/2}$ is the friction velocity. The $Re_b = 20{,}000$ and reference DNS data[35] have $Re_\tau = 543.5$ and $C_{f,DNS} = 0.00591$. $N_x$, $N_y$, and $N_z$ denote the number of grid points in the x-, y-, and z-directions, respectively, while $\Delta x$, $\Delta y$, and $\Delta z$ represent the corresponding grid spacings. Specifically, $\Delta y_c$ denotes the grid spacing at the channel centerline, and $\Delta y_w$ is the height of the first off-wall layer. The velocity is 1.0m/s, eddy viscosity is 1E-4.

Figure 3 illustrates the computational domain, which measures 12.8 m × 2 m × 4.8 m and simulates turbulent channel flow at a friction Reynolds number of 544. A grid independence study was conducted using three mesh resolutions, with total cell counts ranging from approximately 1 million to 10.6 million. Figure 4 presents the instantaneous streamwise velocity field obtained with the medium grid resolution, clearly demonstrating the resolved turbulent structures.

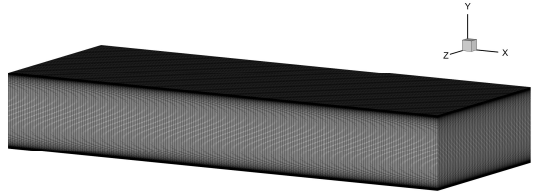

**Fig. 3. Medium grid ($N_x \times N_y \times N_z$ =192 × 128 × 128)**

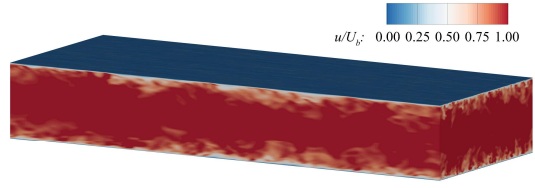

**Fig. 4. Instantaneous streamwise velocity**

Figure 5 presents the four panels displaying the three components and magnitude of the Liutex vector field. The Liutex vector represents an advanced vortex identification method that provides directional information regarding vortical structures while quantifying the local angular velocity of the rigid rotation component within the fluid motion. This characteristic makes it particularly suitable as a foundation for subgrid-scale (SGS) modeling, as it captures the essential physics of turbulent eddies that require parameterization.

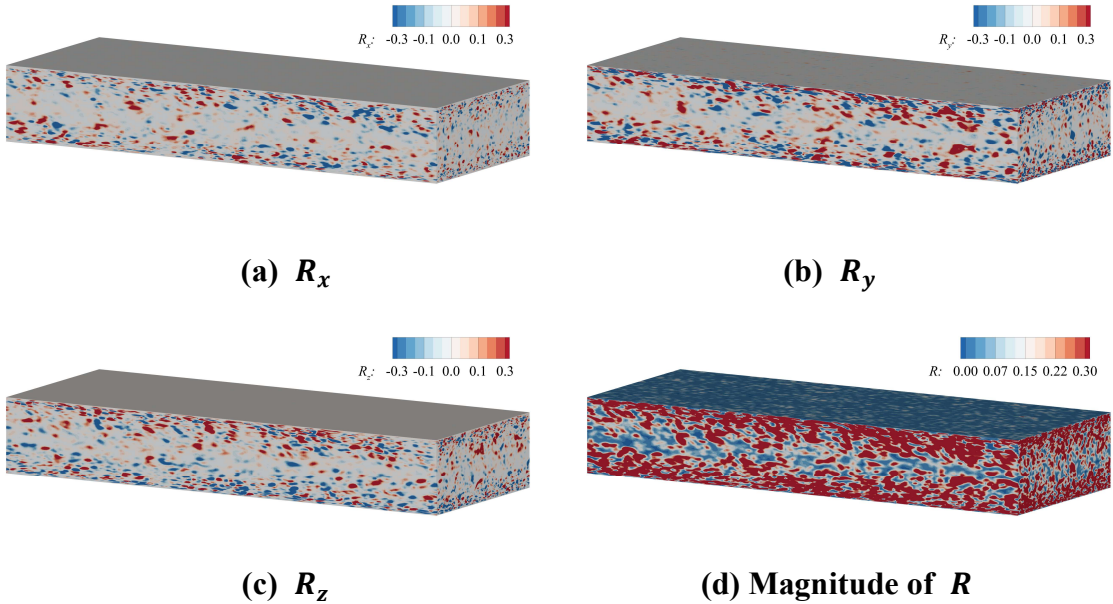

(a) $R_x$          (b) $R_y$

(c) $R_z$          (d) Magnitude of $R$

**Fig. 5. Three components and magnitude of Liutex vector ($R_x$, $R_y$, $R_z$)**

Figure 6 presents a critical comparison among different length scales through dimensionless ratios: $\Delta_{max}/\Delta_x$, $\Delta_{vol}/\Delta_x$ and $\Delta_R/\Delta_x$. The results reveal that $\Delta_R$, the Liutex-based length scale, exhibits significant spatial variation that strongly correlates with the underlying flow structures. In contrast, conventional length scales such as the maximum cell dimension and the cube root of cell volume demonstrate relatively uniform distributions and fail to respond dynamically to local flow conditions. This adaptive behavior constitutes a key advantage of the proposed approach, as the model automatically adjusts to local turbulent characteristics.

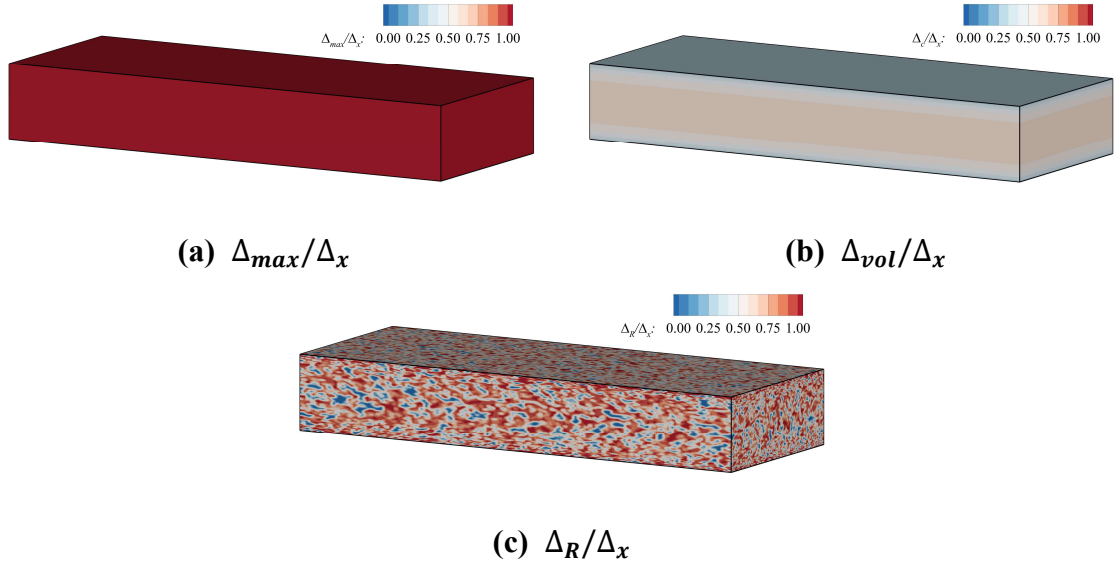

(a) $\Delta_{max}/\Delta_x$  (b) $\Delta_{vol}/\Delta_x$

(c) $\Delta_R/\Delta_x$

Fig. 6. Comparison of different length scales

Figure 7 presents a comparison between the proposed dynamic model coefficient $C_R$ and the classical Germano-Lilly dynamic model coefficient $C_D$. The visualizations demonstrate that the Germano-Lilly model coefficient (panel a) produces significantly larger magnitudes compared to the $C_R$ coefficient (panel b). This disparity in magnitude has important implications for the predicted eddy viscosity and overall model behavior in LES simulations.

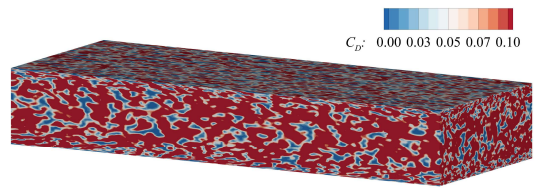

(a) Germano-Lilly model coefficient, $C_D$

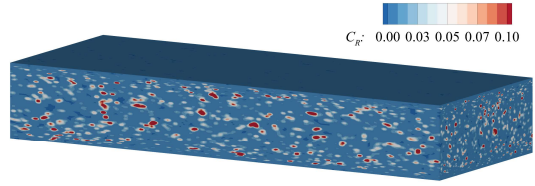

**(b) Dynamic model coefficient based on physics, $C_R$**
**Fig. 7. Comparation of different dynamic model coefficients**

Figures 8 shows the model performance through Reynolds shear stress predictions from different SGS models compared against DNS reference data. The proposed model (shown in red) demonstrates superior performance relative to both the Dynamic Smagorinsky model and Ding et al.[11] model. As the grid resolution transitions from coarse to medium, the differences among various SGS models diminish; however, the proposed model maintains a slight performance advantage, demonstrating consistent accuracy across different grid resolutions—an essential characteristic for practical LES applications.

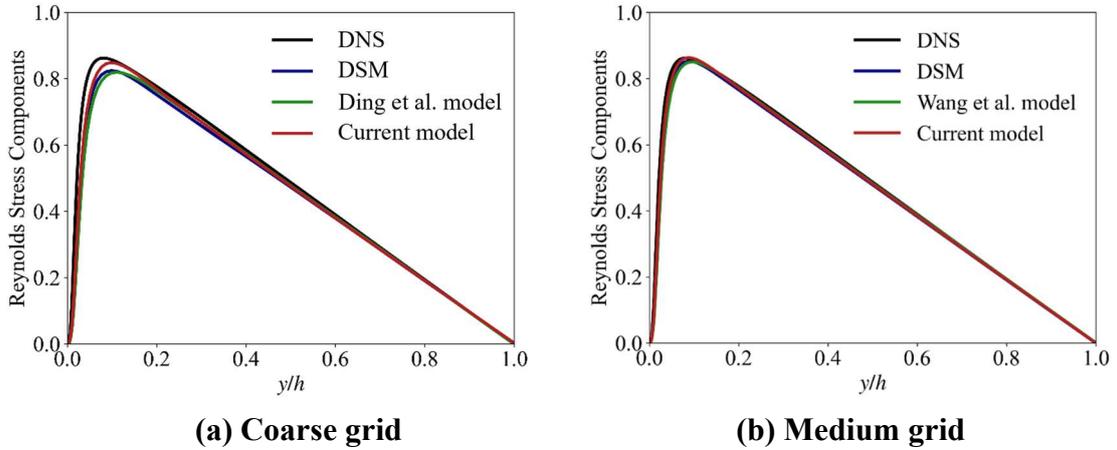

**(a) Coarse grid**  **(b) Medium grid**

**Fig. 8. Model performance through Reynolds shear stress predictions from different SGS models compared against DNS reference data**

Figure 9 shows both normal and shear stress components. For the normal stress (Figure 9(a)), the proposed model captures the near-wall peak more accurately and reproduces the correct decay rate toward the centerline. Similarly, for the shear stress (Figure 9(b)), the model exhibits improved predictions, particularly in the near-wall region where accurate resolution is most challenging. These results confirm that the proposed model provides reliable predictions for the critical Reynolds stress components in wall-bounded turbulent flows.

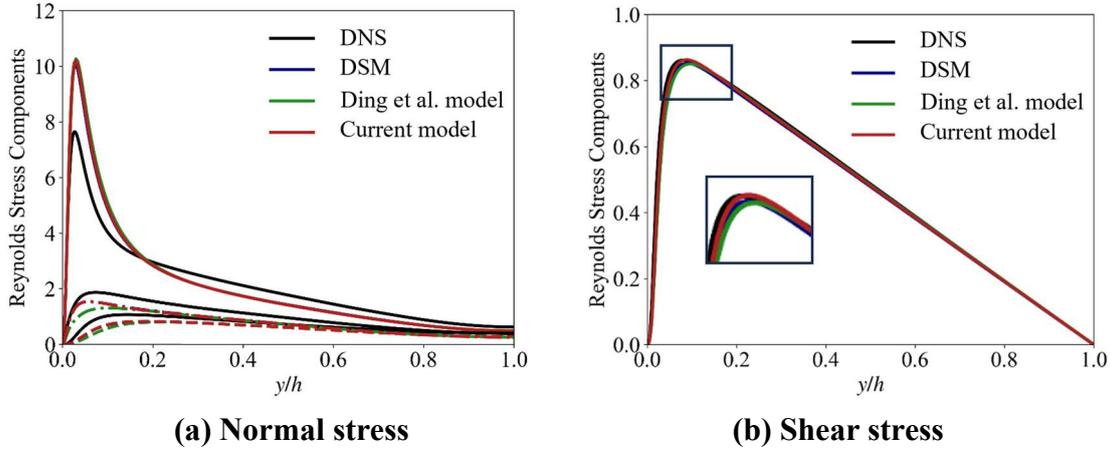

(a) Normal stress　　　　　　　　(b) Shear stress

**Fig. 9. Reynolds normal and shear stress components on the medium grid**

Computational efficiency represents a critical consideration for the practical implementation of any SGS model. To evaluate the computational efficiency of different SGS models, simulations were conducted on the same mesh consisting of 3.13 million cells. The total physical time simulated was 1000 seconds with a fixed time step. All computations were performed on the Texas Advanced Computing Center (TACC) using 128 CPU cores. Tables 1 and 2 present a comparative analysis of computational performance evaluated using 128 processor cores. The proposed model employing the $C_R$ coefficient demonstrates superior computational efficiency compared to Ding et al.[11] model with the Germano-Lilly coefficient, exhibiting only 4.21% overhead relative to the baseline simulation, in contrast to the 6.89% overhead incurred by the latter approach. Furthermore, substantial performance improvement potential exists through algorithmic optimization. Specifically, integrating the $C_R$ calculation directly within the SGS computation framework, rather than employing a separate recalculation procedure, is expected to yield additional efficiency gains. The second table benchmarks the proposed model against established classical SGS formulations, revealing competitive computational costs. Notably, the present model achieves faster execution than the Dynamic Smagorinsky model coupled with the Germano-Lilly coefficient, demonstrating that enhanced physical fidelity need not compromise computational efficiency.

Table 1: Computational speed of different SGS models (128 cores)

| SGS models | Ding et al.[11] model | Ding et al.[11] model | Current model | |
|---|---|---|---|---|
| Model coefficient | 0.19 | Germano-Lilly | $C_R$ | |
| Length scale | $\sqrt[3]{\Delta_x \Delta_y \Delta_z}$ | $\sqrt[3]{\Delta_x \Delta_y \Delta_z}$ | $\sqrt[3]{\Delta_x \Delta_y \Delta_z}$ | $\Delta_R$ |
| Time cost (%) | Reference | +6.89% | +1.66% | +4.21% |

Table 2: Computational speed of different SGS models (128 cores)

| SGS models | Smagorinsky model | Smagorinsky model | WALE |
|---|---|---|---|
| Model coefficient | 0.11 (Van Driest damping) | Germano-Lilly | $C_w$=0.5 |
| Length scale | $\sqrt[3]{\Delta_x \Delta_y \Delta_z}$ | $\sqrt[3]{\Delta_x \Delta_y \Delta_z}$ | $\sqrt[3]{\Delta_x \Delta_y \Delta_z}$ |
| Time cost (%) | +0.55% | +2.07% | 1.41% |

## 4. Conclusions

This study presents a novel subgrid-scale model grounded in Liutex theory, featuring an innovative dynamic model coefficient formulation and a physics-based length scale definition. The model has been rigorously validated using canonical wall-bounded turbulent channel flow at a friction Reynolds number of 544. Comprehensive investigations of the underlying physical mechanisms and Reynolds stress prediction capabilities have been conducted. Several conclusions are drawn from this work:

(1) The proposed SGS model is physically well grounded and highly interpretable. The characteristic velocity scale is defined by the magnitude of the Liutex vector, which directly quantifies the local vortical intensity. The length-scale formulation is guided by the fundamental physical principle that turbulent diffusion predominantly occurs in directions perpendicular to the local rotation axis. Moreover, the dynamic model coefficient is evaluated locally based on the relative strength of vortical structures, enabling a physically consistent and adaptive response to local flow features.

(2) The model exhibits favorable predictive accuracy and computational efficiency. In comparison with the Dynamic Smagorinsky model and the original formulation proposed by Ding et al.[11], the present model provides improved predictions of Reynolds stresses. Importantly, its performance remains stable across different grid resolutions, which is a crucial requirement for practical large-eddy simulation (LES) applications. Furthermore, the computational overhead remains competitive with established SGS models, indicating that enhanced physical fidelity has been achieved without compromising computational efficiency.

Future work will focus on systematic comparisons with a wider range of state-of-the-art SGS models to further benchmark the proposed approach. More importantly, validations involving complex engineering flows with realistic geometries and flow conditions are necessary to assess the model's robustness and to demonstrate its applicability to industrial-scale problems.

## Acknowledgments

This work is supported by the National Science Foundation, with Grant No. 2422573. The authors also thankful to TACC for providing computation resources.

## Author declarations

### Conflicts of Interest
The authors have no conflicts to disclose.

## Data availability

The data that support the findings of this study are available from the corresponding author upon reasonable request.